**Title:** Observation of protein thermodynamics in ice by passive millimeter-wave microscopy

**Authors:** Manabu Ishino[1], Akio Kishigami[2*], Hiroyuki Kudo[3], Jongsuck Bae[4], Tatsuo Nozokido[1*]

**Affiliations:**

[1] Graduate School of Science and Engineering for Research, University of Toyama,

3190 Gofuku, Toyama 930-8555, Japan.

[2] Department of Health and Nutrition, Faculty of Home Economics, Gifu Women's University,

80 Taromaru, Gifu 501-2592, Japan.

[3] Division of Information Engineering, Faculty of Engineering, Information and Systems,

University of Tsukuba,

Tennoudai 1-1-1, Tsukuba, 305-8573, Japan.

[4] Department of Physical Science and Engineering, Nagoya Institute of Technology,

Gokiso, Showa-ku, Nagoya 466-8555, Japan.

* Correspondence to: akio@gijodai.ac.jp (A. K.) and nozokido@eng.u-toyama.ac.jp (T. N.)

**Abstract:**

The study of protein functions attributed to the conformation and fluctuation that are ruled by both the amino acid sequence and thermodynamics, requires thermodynamic quantities given by calorimetry using thermometric techniques. The increased need for protein function in different applications requires improvements of measurement systems assessing protein thermodynamics to handle many kinds of samples quickly. We have developed a passive millimeter-wave microscope that allows near-field imaging of thermal radiation, even in the low temperature range under room temperature where passive infrared imaging systems are ineffective. This advantage of our microscope system in combination with low thermal emission property of water ice in the millimeter-wave region enables the characterization of the thermal radiation from the proteins themselves in aqueous solution at a temperature range low enough to freeze water and to trap conformation intermediates in the proteins. Experiments performed at a millimeter-wave frequency of 50 GHz in a temperature range from 130 K to 270 K for a 20 % bovine serum albumin (BSA) aqueous solution showed a displacement between two conformational states of BSA at a temperature of approximately 190 K as a boundary. Our microscope system using this freeze-trapping method is expected to provide noninvasive thermal images to enable novel high-throughput calorimetry useful for the analysis of protein functions.

# 1. Introduction:

Proteins function with high efficiency, specificity, and accuracy as a building block of life. Their flexible structure thermally fluctuates with many motions over broad timescales ranging from milliseconds to picoseconds. Each protein function provides evidence that motions in its protein fluctuation are not just random but are related to biological functionality, with routine exchanges among structural states [1]. Many studies by time-resolved crystallographic methods, nuclear magnetic resonance spectroscopy, and a range of other biophysical techniques and computational methods have demonstrated an intricate view of the multiple conformations of proteins to relate their functions [1, 2, 3]. Many biological processes such as protein folding and conformational changes leading to physiological functions like catalysis and signaling occur on the slow timescale from micro- to milliseconds at physiological temperature, indicating that the collective motions of large-scale groups of amino acids in protein that correspond to the slow timescale are important in protein functions [1, 4, 5]. Because all motions are activated by thermal energy, at sufficiently low temperature — generally below 200 K for most of the collective motions — insufficiency of thermal energy ceases individual motions and prevents protein functions [1, 2]. This means that a sufficiently low temperature to limit the collective motions consisting of conformational states traps protein fluctuations in some intermediates that are interconverted by jumping over energy barriers between conformational states in the protein energy landscape. Studies by spectroscopy combined with the freeze-trapping method have developed the concepts to describe proteins in the relation between function and conformation transition that is revealed by rising temperature [2, 5, 6, 7].

Understanding protein functions requires information on thermodynamics described by the free energy including enthalpy and entropy to figure out dynamics in the interconversion of protein conformations [1, 2]. Calorimetric experiments for proteins have provided thermodynamic parameters [8]. Two major calorimetric techniques that directly measure heat change are differential scanning calorimetry (DSC), which measures sample heat capacity ($C_p$) with respect to a reference sample as a function of temperature, and isothermal titration calorimetry (ITC), which measures the heat change during titration [8, 9]. These methods require substantial time for either referencing (in DSC) or titration (in ITC) and need to be in contact with an object in their temperature measurements to achieve heat conduction and a thermal equilibrium state. Instead of measuring heat directly, calorimetry is also achieved using the temperature dependence of the equilibrium constants measured by appropriate methods, according to the van't Hoff relations to obtain enthalpy and heat capacity changes [8, 9]. Although the van't Hoff equation is integrated between two temperatures under the assumption that the reaction enthalpy is constant, the reaction enthalpy mostly changes in practice, resulting in an approximate integration [10]. In spite of these disadvantages, estimation of thermodynamic parameters by calorimetry has contributed to the interpretation of protein dynamics for their biological relevance [4-9]. These contributions of thermodynamic measurements push forward the demand for more improvements in calorimetry through extensions of mature technology.

Radiation thermography is another thermometric technique that noninvasively images an object by detecting electromagnetic radiation thermally emitted from the object over a short time period, and relates the energy of the thermal radiation to the object temperature using the emissivity ($\varepsilon$) of the object, which is equal to the object's absorptance ($\alpha$), under Planck's law [11]. Conventional thermographs detect thermal radiation in the infrared region for imaging. The power of the infrared radiation decreases drastically under 300 K as shown by the red line in Fig. 1, which shows the spectral radiance at 10 μm wavelength as a function of blackbody ($\varepsilon = 1$) temperature calculated by Planck's law [12, 13]. Because of this power decrease, passive infrared imaging instruments show that approximately 230 K is their limit for low temperature operation. However, the intensity of the thermal radiation in the millimeter-wave region is proportional to the temperature down to a few K, as shown with the blue line in Fig. 1, which is the spectral radiance at a millimeter-wave wavelength of 6 mm. Therefore, thermography using millimeter-wave radiation is superior below room temperature to that using infrared radiation. Thermography using millimeter-wave radiation is insufficient in its wavelength because the spatial resolution achievable with radiometric imaging system using conventional optics is limited to the wavelength of the operating frequency because of the diffraction effect [14], causing necessity for preparing a large amount of sample. We have overcome this problem by developing a tapered near-field slit probe that allows for passive thermal imaging in the millimeter-wave region with subwavelength spatial resolution below the diffraction limit [15, 16]. Our passive millimeter-wave microscope performed thermal imaging in the temperature range from room temperature down to 160 K [16]. Because water ice emits negligible thermal radiation in the millimeter-wave region [17], when our microscope system scans a sample in ice, the thermal radiation from samples other than solid water should be detected. Our thermometric technique described in the following has the advantage of enabling the observation of proteins in frozen aqueous solution in a low temperature range that terminates the slow timescale motions in proteins.

## 2. Materials and Methods

We chose an aqueous solution of bovine serum albumin (BSA) as the target sample in this experiment because BSA is easily obtained and prepared. The properties of BSA have also been characterized by many biophysical studies including calorimetry by other methods [18-21]. BSA (initial fraction by heat shock, fraction V, above 98 % purity by electrophoresis) was purchased from Sigma-Aldrich Japan, (Tokyo, Japan). Twenty (wt/vol) percent aqueous solution of BSA was prepared with an appropriate amount of purified water.

A passive millimeter-wave microscope system operating at 50 GHz (6 mm wavelength) was used to observe thermodynamic aspects in thermal radiation emitted from protein samples in ice. Figure 2 is a schematic drawing of our experimental setup. As a near-field sensor, we used a tapered slit probe with a 4-section quarter-wave transformer newly developed for this work. The height of the WR-19 metallic rectangular waveguide for U-band (40-60 GHz) frequencies reduced to form a slit-like aperture at the probe tip. This type of probe can be operated with high transmission efficiency [15]. The long and short dimensions of the slit aperture are 4.8 mm and 150 μm, respectively The probe tip was placed in close proximity to the sample. An aluminum sample holder was filled with pure water or a 20 % BSA solution and was placed on a copper sample stage, which was temperature-controlled using liquid nitrogen ($N_2$) as a coolant and an electric heater in combination with a temperature controller operating in on/off or proportional-integral-derivative mode. The temperature of the stage was measured using a thermocouple built into it just beneath the sample. The sample surface was covered with an 11-µm-thick polyvinylidene chloride film. The probe-to-sample separation was adjusted to 20 µm using a piezoelectric transducer and a laser displacement sensor during measurements.

The millimeter-wave radiation thermally emitted from the sample was converted via the slit aperture into propagating waves in the probe waveguide, which were then detected by the receiver shown in Fig. 3. The experimental setup shown in Fig. 2 and part of the receiver shown in Fig. 3 were sited in a chamber made of acrylic resin. The chamber was filled with dry $N_2$ gas evaporated from liquid $N_2$ to prevent the sample surface from becoming frosted. The measurement system was installed on an optical bench.

Fig. 3 shows a block diagram of the radiometric receiver, which operates in a Dicke-switched mode [22]. The input to the receiver was switched by a PIN diode switch between the probe and a termination with a repetition frequency of 1 kHz. The output port of the switch was connected to a low-noise amplifier (LNA) via two isolators connected in series. The output of the LNA was down-converted to an intermediate frequency (IF) signal with a center frequency of 1 GHz by a single-sideband mixer. The IF bandwidth was 0.4 GHz. The IF signal was amplified and then detected using a diode detector. The voltage difference between the outputs from the detector when the probe and the termination were connected was measured using a lock-in amplifier. The cyan-colored components of the receiver shown in Fig. 3 were sited inside the chamber. The time constant of the lock-in amplifier was 3 seconds.

## 3. Theory

The signal processing method for the output from the lock-in amplifier to extract emissivity information of the BSA sample is described in the following. The incident electromagnetic energy on an object is divided into whether it is absorbed by the object (absorptance: $\alpha$ ), reflected from the object (reflectance: $r$), and transmitted through the object (transmittance: $t$). The relation between these quantities holds in the forms:

$$\alpha + r + t = 1 \tag{1}$$

and

$$\varepsilon = \alpha = 1 - (r + t), \tag{2}$$

where $\varepsilon$ is the emissivity of the object. This emissivity relates the measured energy to the physical temperature of the object in radiation thermography.

Figure 4 shows how the brightness temperature of the noise signal coming from the probe is calculated. The probe tip is placed in close proximity to the sample surface. The sample is characterized by the physical parameters $T_s$, $r$, $t$, and $\varepsilon$, where $T_s$ is the physical temperature. The sum of the parameters $r$, $t$, and $\varepsilon$ equals unity because of the energy conservation law and Kirchhoff's law as shown in Eq. (2). The brightness temperature of the noise signal coming from the probe $T_{pr}$ can be calculated as

$$T_{pr} = \gamma T_r + (1-\gamma)\{\varepsilon T_s + (1-\varepsilon)T_{am}\}, \tag{3}$$

where $\gamma$, $T_r$, and $T_{am}$ are the reflectance of the probe, the physical temperature of the receiver, and the ambient temperature that expresses the influence of ambient thermal radiation, respectively [14, 16]. The first term in the right hand side of Eq. (3) shows the reflection of the thermal radiation from the receiver that is incident on the probe. The second term on the right hand side of Eq. (3) shows the transmission of the thermal radiation emitted from the sample plus the thermal radiation at ambient temperature that is both reflected at the sample surface and transmitted through the sample via the slit aperture to the probe waveguide.

The output signals of the receiver shown in Fig. 3 when the probe and the termination are connected $S_{pr}$ and $S_{term}$ are written as

$$S_{pr} \propto (T_{pr} + T_{sys})kGB \tag{4}$$

and

$$S_{term} \propto (T_r + T_{sys})kGB, \tag{5}$$

respectively. $k$, $G$, $B$, and $T_{sys}$ are Boltzmann's constant, the gain of the receiver, the bandwidth of the receiver, and the system noise temperature of the receiver, respectively. The output from the lock-in amplifier $S$ is the difference between the above two signals

$$S \propto S_{pr} - S_{term} = -(1-\gamma)T_r + (1-\gamma)\{\varepsilon T_s + (1-\varepsilon)T_{am}\}. \tag{6}$$

The output signal that is free from the effect of the mismatch between the probe and the sample $S_f$ can be obtained by dividing $S$ by the transmittance of the probe $1-\gamma$ as follows:

$$S_f = \frac{S}{1-\gamma} \propto -T_r + \varepsilon T_s + (1-\varepsilon)T_{am}. \tag{7}$$

Because the emissivity of water ice is negligible [17], the output signals that are free from the effect of the mismatch when water ice and a BSA solution are taken as a sample $S_{f\_BSA}$ and $S_{f\_ice}$ can be written as

$$S_{f\_BSA} = \frac{S_{BSA}}{1-\gamma_{BSA}} \propto -T_r + \varepsilon T_s + (1-\varepsilon)T_{am} \tag{8}$$

and

$$S_{f\_ice} = \frac{S_{ice}}{1-\gamma_{ice}} \propto -T_r + T_{am}, \tag{9}$$

respectively. $\varepsilon$ in Eq. (8) is the emissivity of the BSA sample. $S_{BSA}$ and $S_{ice}$ are the outputs from the lock-in amplifier when the BSA solution and water ice are used as a sample, respectively. $\gamma_{BSA}$ and $\gamma_{ice}$ are the reflectances of the probe when the BSA solution and water ice are used, respectively. The signal intensity when the signal of water ice is subtracted from that of BSA can be calculated as

$$S_{f\_BSA} - S_{f\_ice} \propto \varepsilon(T_s - T_{am}). \tag{10}$$

Equation (10) indicates that when $T_{am}$ is constant against the variation of $T_s$ or proportional to $T_s$, the slope of the signal $S_{f\_BSA} - S_{f\_ice}$ against the sample temperature provides an estimate of the BSA emissivity.

## 4. Results and Discussion

Figure 5A shows the voltage outputs from the lock-in amplifier, which are the final output signals from the measurement system, as functions of the temperature of the sample stage, when water and a 20% aqueous solution of BSA were placed in an aluminum sample holder, respectively. On the vertical axis in Fig. 5A, the positive value means that the signal received from the probe is greater than that from the non-reflective termination kept at environmental temperature. While the output of water increased proportionally to the temperature of the sample stage, the output of the BSA solution showed a quite different pattern. Before processing the outputs in order to extract emissivity information of BSA, the output signals free from the effect of the mismatch between the probe and each sample should be calculated as shown in Eqs (7)-(9) because the difference in mismatch causes difference in measurement sensitivity, which should be compensated.

Figure 5B shows reflectances of the probe as functions of the sample stage temperature when water and the BSA solution were taken as a sample, respectively. The reflectance value of zero means there is no mismatch. When measuring the reflectance of the probe, the cyan and yellow colored components of the receiver shown in Fig. 3 were replaced with a circulator. A signal from an amplitude-modulated noise source was amplified, and then fed to the probe via the circulator. The reflected signal from the probe passing through the circulator was recorded with the receiver. Total reflection was calibrated using a metal plate instead of the probe. Because the reflectance was measured using an output from a noise source as an input to the probe and the reflected signal from the probe was measured using a part of the receiver as a detector, the reflectance of the probe for the thermally emitted noise-like signal was measured accurately. The signal intensity when the signal of water ice is subtracted from that of BSA as described in Eq. (10) is shown in Fig. 5C. This result indicates that the signal intensity increases proportionally to the temperature from 130 K to 170 K and from 230 K to 270 K, but decreases within the range 170 K to 230 K. Because the output of water ice is proportional to the sample temperature as shown in Fig. 5A and the reflectance is nearly constant in Fig. 5B, the slope of the signal against the temperature in Fig. 5C provides an estimate of the BSA emissivity. We found that quantification of the BSA emissivity is possible when the data obtained from a sample with a known emissivity are available.

The simplest model that explains our results described above is a displacement from an equilibrium state to another state that has a different emissivity. One equilibrium state below 170 K has an emissivity calculated by linear approximation that is different from that of another state over 230 K in Fig. 5C. The differentiation of the signal in Fig. 5C with respect to the temperature gives the result in Fig. 5D, showing a parabola in the temperature range from 170 K to 230 K with the minimum value at approximately 190 K. This minimum value corresponds to the reported values in the literature for the glass transition temperature in the case of BSA aqueous solutions [18-21]. Previous studies for BSA solutions have suggested that this glass transition originates from the cooperative motion of protein and hydrated water around BSA [18-20]. Because the signal from 170 K to 230 K corresponds to the transition phase between two states, the minimum value indicates the energy barrier or the activation energy for the transition. In addition, the relation in which the slope at the lower temperature range is nearly half that at the higher range in Fig. 5C means that the $C_p$ of the conformational state over 230 K of BSA is 2-fold larger than that below 170 K because the rate of change in heat emissivity homologous to heat absorptance under thermal equilibrium conditions is equivalent to the rate of change in $C_p$ after the transition. This result agrees with the study for BSA solution by DSC reporting two $C_p$s of 5 and 10 kJK$^{-1}$mol$^{-1}$ [19].

## 5. Conclusion

Our experiments show thermodynamics in the transition of structural states in protein as the transition in emissivity. Advantages of our thermometric method are that the measurement is noninvasive, information from ice is free, and the system can provide two-dimensional thermal images [16]. Our system is expected to overcome common problems in protein calorimetry. These advantages offer potential for the development of thermographic instruments to handle many kinds of samples quickly. Array calorimetry using higher-throughput instruments like ours, that can scan over arrayed wells that hold different samples made under various conditions, is required in the assessment of the relation between protein structure and function; for instance, in structural biology-based drug discovery by characterization of ligand binding by enthalpy [23-25].

**Acknowledgements:**

This work was supported in part by JSPS KAKENHI Grant Nos. 16K14684, 15H04016, and 18H01450.

**References:**

1. Henzler-Wildman K and Kern D (2007) Dynamic personalities of proteins. *Nature* **450**, 964-972.

2. Lewandowski JR, Halse ME, Blackledge M and Emsley L (2015) Direct observation of hierarchical protein dynamics. *Science* **348**, 578-581.

3. Papaleo E, Saladino G, Lambrughi M, Lindorff-Larsen K, Gervasio FL and Nussinov R (2016) The role of protein loops and linkers in conformational dynamics and allostery. *Chem. Rev.* **116**, 6391–6423.

4. Ansari A, Berendzen J, Bowne SF, Frauenfelder H, Iben IET, Sauke TB, Shyamsunder E and Young RD (1985) Protein states and proteinquakes. *Proc. Natl. Acad. Sci. U.S.A.* **82**, 5000-5004.

5. Frauenfelder H, Sligar SG and Wolynes PG (1991) The energy landscapes and motions of proteins. *Science* **254**, 1598-1603.

6. Wald G, Durell J and St. George RCC (1950) The light reaction in the bleaching of rhodopsin. *Science* **111**, 179-181.

7. Kim TH, Mehrabi P, Ren Z, Sljoka A, Ing C, Bezginov A, Ye L, Pomes R, Prosser RS, and Pai EF (2017) The role of dimer asymmetry and protomer dynamics in enzyme catalysis. *Science* **355**, 262.

8. Cooper A, Johnsonb CM, Lakey JH and Nollmanna M (2001) Heat does not come in different colours: Entropy-enthalpy compensation, free energy windows, quantum confinement, pressure perturbation calorimetry, solvation and the multiple causes of heat capacity effects in biomolecular interactions. *Biophys. Chem.* **93**, 215-230.

9. Prabhu NV and Sharp KA (2005) Heat capacity in proteins. *Annu. Rev. Phys. Chem.* **56**, 521-548.

10. Naghibi H, Tamura A and Sturtevant JM (1995) Significant discrepancies between van't Hoff and calorimetric enthalpies. *Proc. Natl. Acad. Sci. U.S.A.* **92**, 5597-5599.

11. Planck M (1901) Über das gesetz der energieverteilung im normalspektrum. *Ann. Physik* **309**, 553-563.

12. Ulaby FT, Moore RK and Fung AK (1981) "Radiometry" in *Microwave remote sensing: Active and passive. Vol. 1 - Microwave remote sensing fundamentals and radiometry*. (Artech House, Massachusetts), chap. 4.

13. Jones A.C, O'Callahan BT, Yang HU and Raschke MB (2013) The thermal near-field: Coherence, spectroscopy, heat-transfer, and optical forces. *Prog. Surf. Sci.* **88**, 349-392.

14. Appleby R and Anderton RN (2007) Millimeter-wave and submillimeter-wave imaging for security and surveillance. *Proc. IEEE* **95**, 1683-1690.

15. Nozokido T, Bae J and Mizuno K (2001) Scanning near-field millimeter-wave microscopy using a metal slit as a scanning probe, *IEEE Trans. Microw. Theory Tech.* **49**, 491-499.

16. Nozokido T, Ishino M, Kudo H and Bae J (2013) Near-field imaging of thermal radiation at low temperatures by passive millimeter-wave microscopy. *Rev. Sci. Instrum.* **84**, 036103-1-3.

17. Iwabuchi H and Yang P (2011) Temperature dependence of ice optical constants: Implications for simulating the single-scattering properties of cold ice clouds. *J. Quant. Spectrosc. Radiat. Transfer* **112**, 2520-2525.

18. Kawai K, Suzuki T and Oguni M (2006) Low-temperature glass transitions of quenched and annealed bovine serum albumin aqueous solutions. *Biophys. J.* **90**, 3732–3738.

19. Shinyashiki N, Yamamoto W, Yokoyama A, Yoshinari T, Yagihara S, Kita R, Ngai KL and Capaccioli S (2009) Glass transitions in aqueous solutions of protein (bovine serum albumin). *J. Phys. Chem. B* **113**, 14448-14456.

20. Panagopoulou A, Kyritsis A, Serra RS, Ribelles JLG, Shinyashiki N and Pissis P (2011) Glass transition and dynamics in BSA-water mixtures over wide ranges of composition studied by thermal and dielectric techniques. *Biochimica et Biophysica Acta* **1814**, 1984-1996.

21. Frontzek AV, Strokov SV, Embs JP and Lushnikov SG (2014) Does a dry protein undergo a glass transition? *J. Phys. Chem. B* **118**, 2796−2802.

22. Ulaby FT, Moore RK and Fung AK (1981) "Radiometer systems" in *Microwave remote sensing: Active and passive. Vol. 1 - Microwave remote sensing fundamentals and radiometry*. (Artech House, Massachusetts), chap. 6.

23. Torres FE, Recht MI, Coyle JE, Bruce RH and Williams G (2010) Higher throughput calorimetry: Opportunities, approaches and challenges. *Curr Opin Struct Biol.* **20**, 598-605.

24. Baron R and McCammon JA (2013) Molecular recognition and ligand association. *Annu. Rev. Phys. Chem.* **64**, 151-175.

25. Renaud J-P, Chung C, Danielson UH, Egner U, Hennig M, Hubbard RE, Nar H (2016) Biophysics in drug discovery : Impact, challenges and opportunities. *Nature reviews Drug discovery* **15**, 679-698.

**Figure Captions**

**Fig. 1**

Spectral radiances at 10 μm and 6 mm wavelengths as functions of blackbody temperature. The red line shows radiance at 10 μm wavelength in the infrared region, calculated by Planck's law. The blue line shows radiance at 6 mm wavelength in the millimeter-wave region. The dotted line indicates room temperature of 300 K.

**Fig. 2**

Experimental setup for passive millimeter-wave microscopy.

**Fig. 3**

Block diagram of the radiometric receiver.

**Fig. 4**

Brightness temperature of the noise signal coming from the probe.

**Fig. 5**

Experimental results.

(A) Voltage outputs from lock-in amplifier when the temperature of the sample stage was varied. Blue and red circles indicate the outputs when water and a 20 % BSA solution were used as a sample, respectively. (B) Reflectances of the probe as functions of the temperature of the sample stage. Blue and red circles indicate the reflectances when water and the BSA solution were used as a sample, respectively. (C) Signal intensity when the signal of water is subtracted from that of BSA. (D) Differentiation of the signal intensity of BSA with respect to the temperature. After averaging for each data with two data in both sides in C, differentiation was performed.

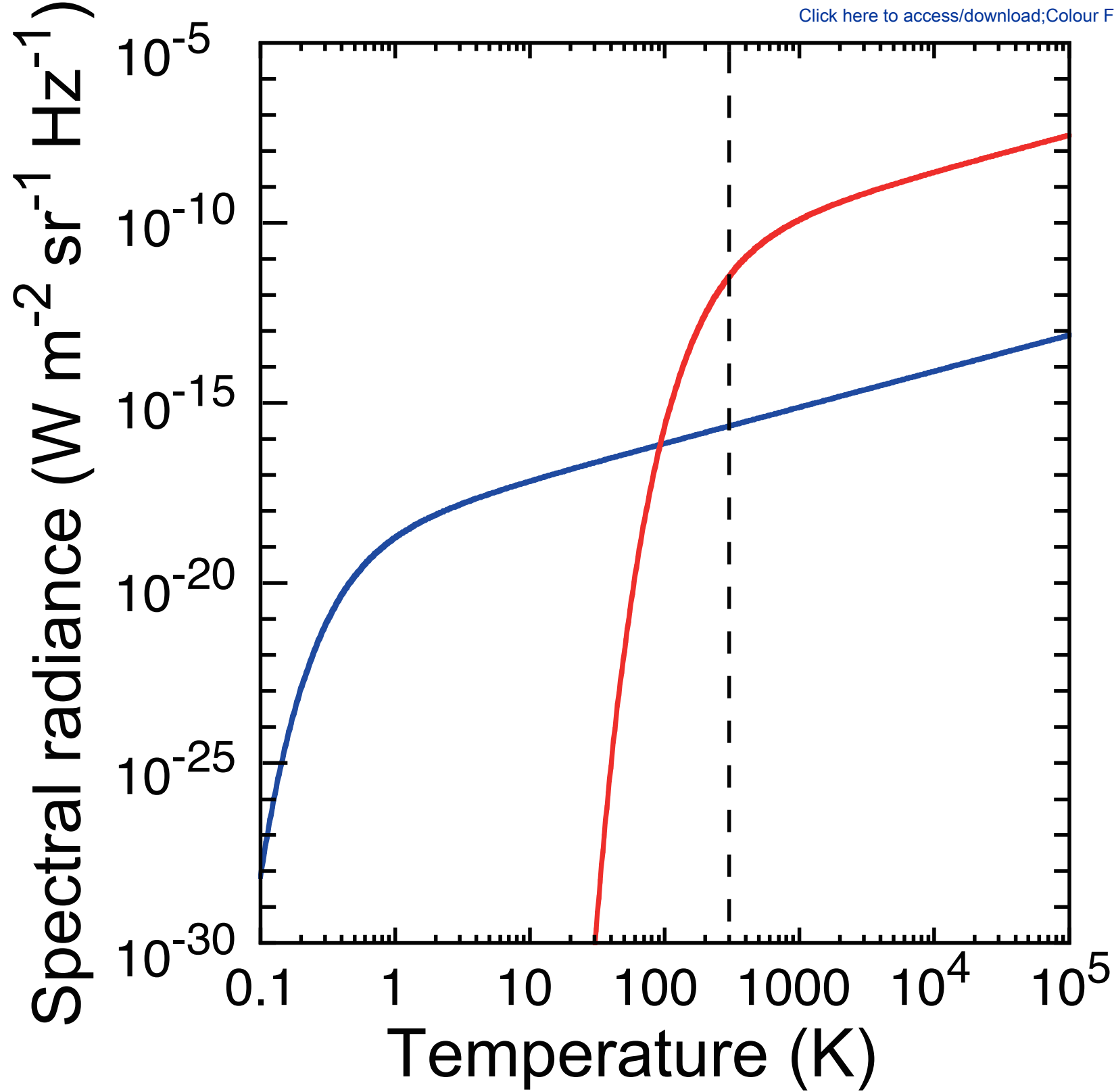

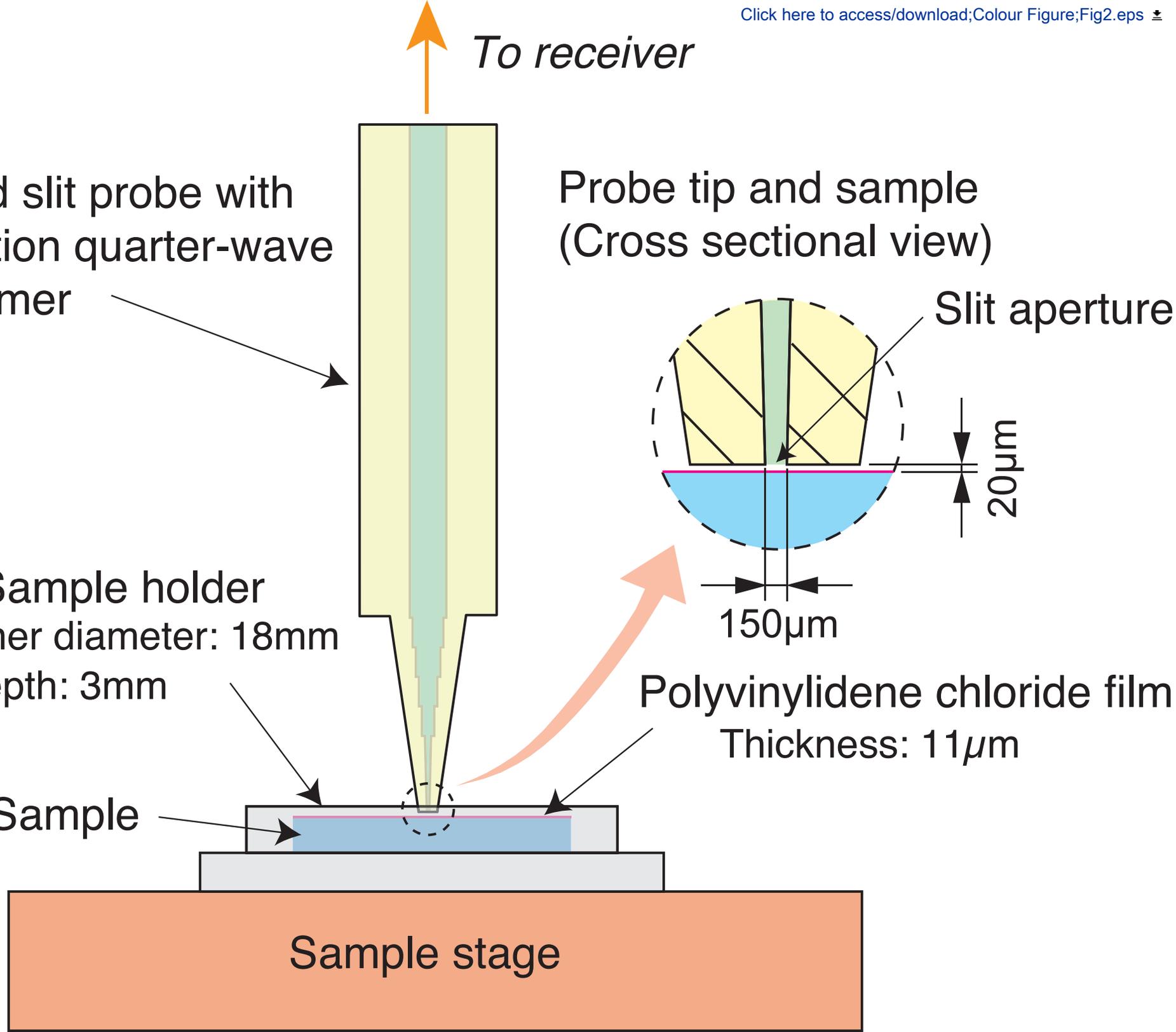
Figure 2

Figure 3

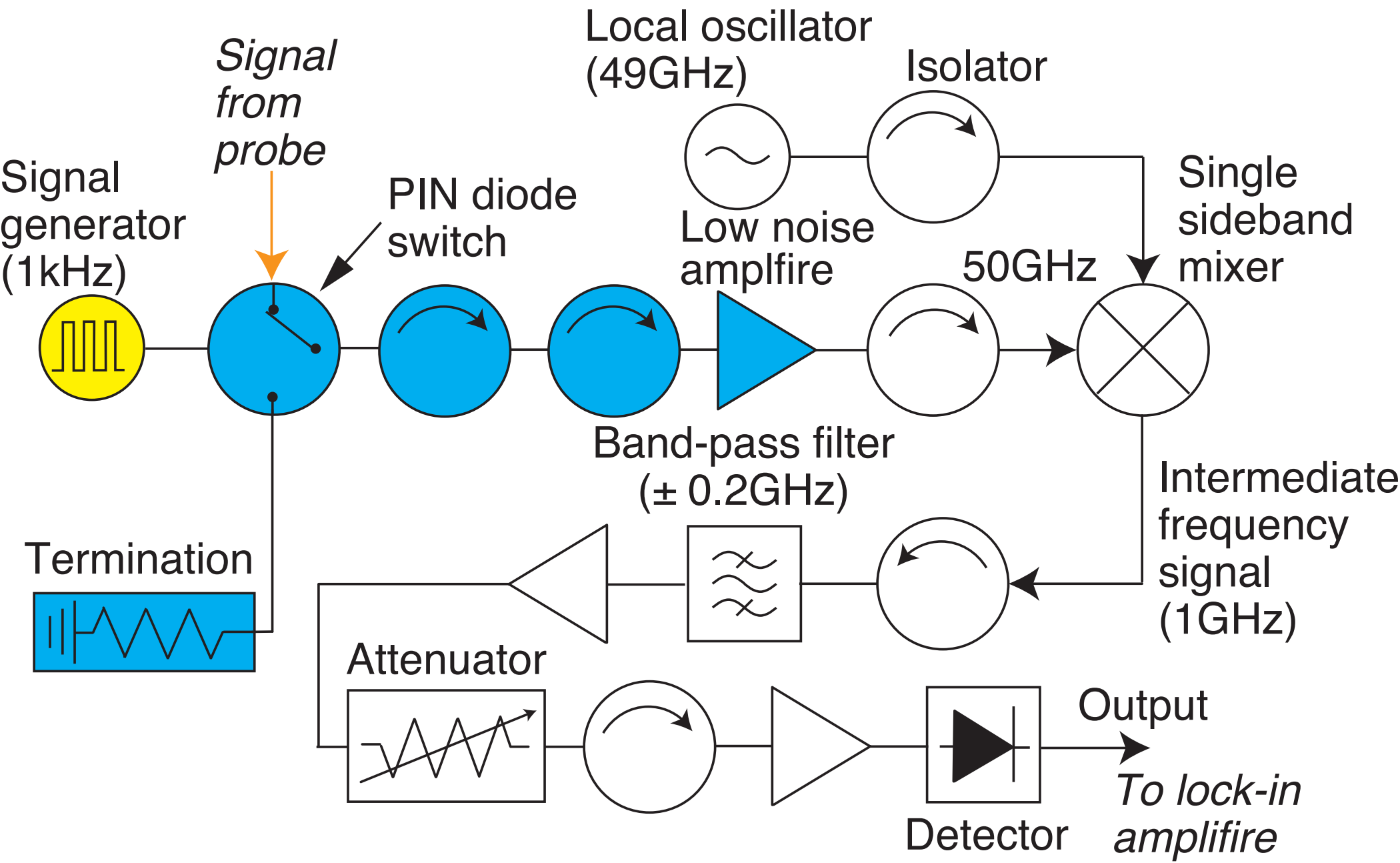

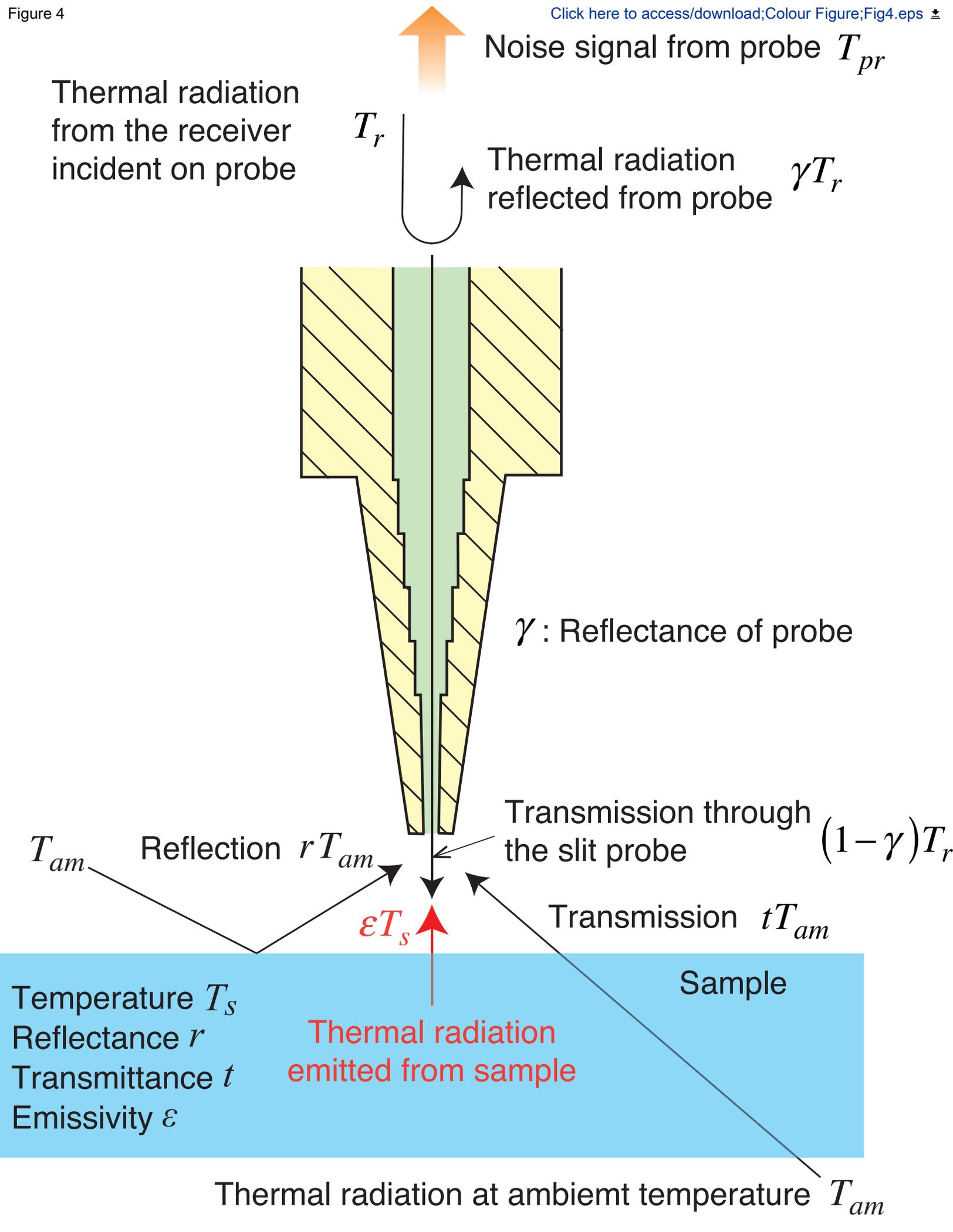

Figure 4

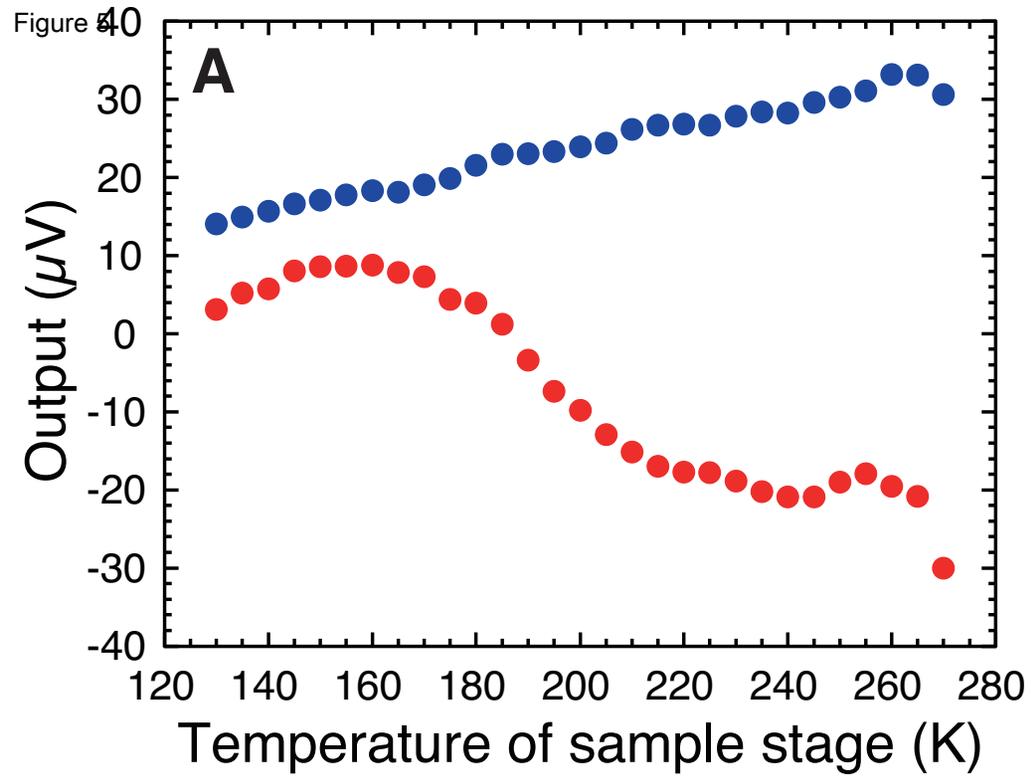
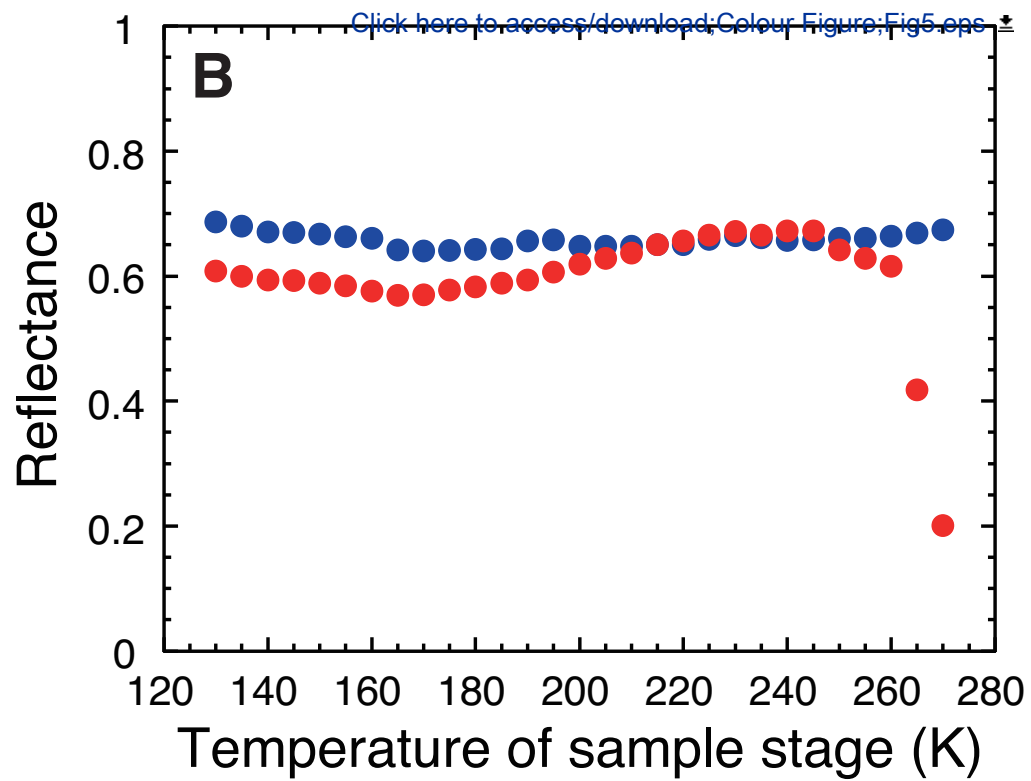
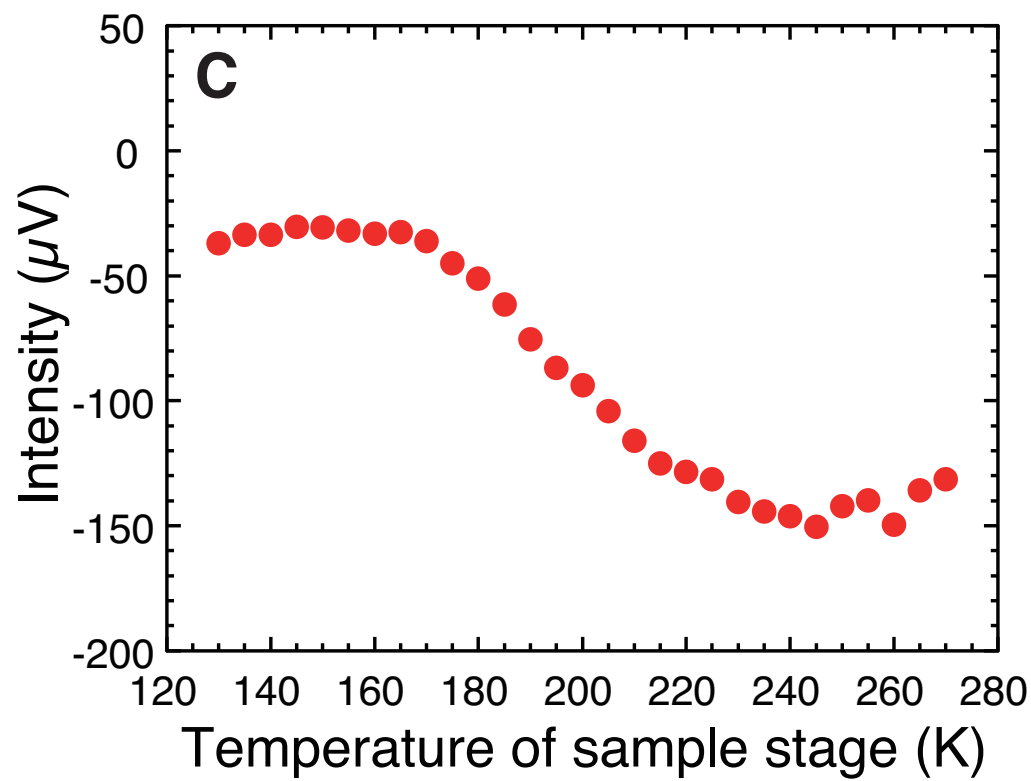
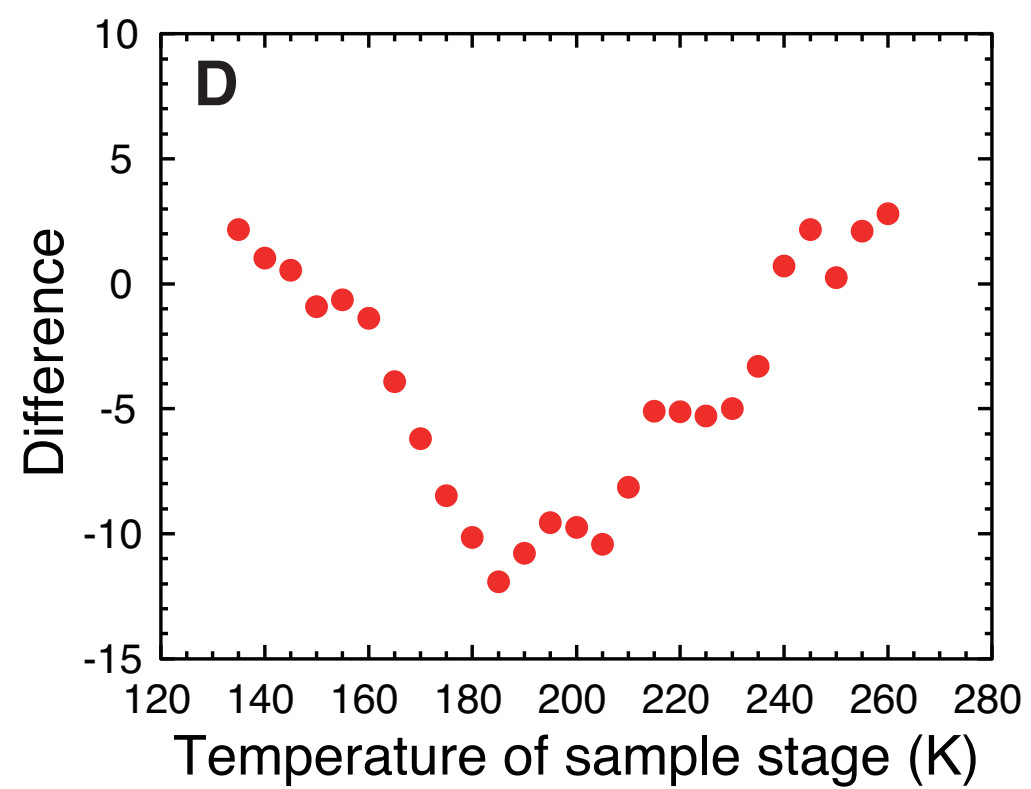